
\centerline {\bf Uncertainty Relation at Finite Temperature}
\vskip .5cm
\centerline{\bf B. L. Hu $^{\ast}$}
\affil{Department of Physics, University of Maryland,
College Park, MD 20742, USA}
\vskip 1cm
\centerline{\bf Yuhong Zhang ${\ast}$}
\affil{Biophysics Lab, Center for Biologics Evaluation and Research,
Food and Drug Adminstration, 8800 Rockville Pike, Bethesda, MD 20982, USA}
\vskip .5cm
\centerline{umdpp 93-161}
\vskip .5cm
\abstract  We use the quantum Brownian model to derive the uncertainty
relation for a quantum open system. We examine how the fluctuations of
a quantum system evolve after it is brought in contact with a heat bath
at finite temperature.
We study the decoherence and relaxation processes and use this example
to examine 1) the relation between quantum and thermal fluctuations;
and 2) the conditions
when the two basic postulates of quantum statistical mechanics become valid.


\vskip 4cm
\noindent Submitted to Phys. Rev. Lett., July 30, 1992

\noindent $^{\ast}$ bitnet addresses: hu@umdhep, zhang@umdhep

\body
\endtitlepage
\vfill
\eject

It is a well-known fact in quantum mechanics that a lower bound exists in
the product of the variances of pairs of noncommutative observables.
Taking the coordinate $x$ and momentum $p$ as examples, the Heisenberg
uncertainty principle states that with $(\Delta x)^2=~<x^2>-~<x>^2$,
the uncertainty function
$$
U_0 ^{QM} =(\Delta x)^2(\Delta p)^2 \geq {\hbar^2 \over 4}
{}~~~(T=0,~~ quantum~ mechanics)
                                                              \eqno(1)
$$
The existence of quantum fluctuations is a verified basic physical
phenomenon. The origin of the uncertainty relation can be attributed as
a mathematical property of Fourier analysis [1] which describes quantum
mechanics as a wave theory. Recent years have seen effort in establishing
a stronger relation based on information-theoretical considerations [2][3].

In realistic conditions quantum systems are often prepared and studied
at finite temperatures where thermal fluctuations permeate. At high
temperatures the equipartition theorem of classical statistical mechanics
imparts for each degree of freedom an uncertainty of $kT/2$.
Thus the uncertainty function for a one-dimensional particle
approaches the limit
$$
U_T ^{MB} \approx ({kT\over\Omega})^2
{}~~~(high~T,~~ classical~ statistical~ mechanics)
                                                                \eqno(2)
$$
where $\hbar \Omega$ is the energy of a normal mode with physical frequency
$\Omega$.
This result, obtained by assuming that the system obeys the
Maxwell-Boltzmann distribution, is  usually regarded as the classical limit.
For a system of bosons in equilibrium at temperature $T$, the application
of canonical ensemble gives the result in quantum statistical mechanics
as
$$
U_T ^{BE} = {\hbar^2 \over 4} [ \coth ({{ \hbar \Omega} \over {2 kT}})]^{2}
{}~~~(all~ T,~~quantum~ statistical~ mechanics)
                                                                \eqno(3)
$$
which interpolates between the two results (1) and (2) at $T=0$ and
$T>> \hbar \Omega /k$. This result applies to a system already in
equilibrium at temperature $T$.

Our purpose here is to study the corresponding non-equilibrium problem.
At time $t_0$ we put the system in contact with a heat bath at temperature
$T$ and follow its time evolution. We want to see how the uncertainty
function $U_T(t)$ changes from the initial quantum fluctuation-dominated
condition to a later thermal fluctuation-dominated condition. By comparing
this result with the decoherence studies recently carried out [5] [6],
where two characteristic times---the decoherence time $t_{dec}$ and the
relaxation time $t_{rel}$---are defined, one can apply the physics of
these two processes involved to examine the following three issues :

\noindent 1) the realization of the basic tenets of quantum statistical
mechanics from quantum dynamics;
2) the relation between quantum and thermal fluctuations; and
3) the quantum to classical transition.


Quantum statistical mechanics of a macroscopic system is derived from
the quantum dynamics of its microscopic constituents under two basic
postulates [7]: i) random phase, and ii) equal {\it a priori} probability.
The first condition enables one to assign probability distributions to
a system occupying certain quantum states. It requires the disappearance
of interference terms in the wave function or that the density matrix
of the system be approximately diagonal. The second condition ensures
that the system when put in contact with a large bath equilibrates with
it. We want to examine the processes by which these two conditions are
attained from a more primitive level, starting with the microdynamics of
a system of quantum particles. Specifically, we want to see if there is
a characteristic time  when the phase information is lost (Postulate i)
and another time when the system attains equilibrium with its surrounding
so that all accesible states are equally probable (Postulate ii).

On the second issue, the demarkation of classical, quantum and thermal regimes
are not always clearly noted, their usual definitions or usage oftentimes
are imprecise. We will point out with the aid of the results obtained here
some existing confusions and make their meanings more precise. The relation
between quantum and thermal fluctuations has been studied previously via thermo
field dynamics [8] under equilibrium and stationary conditions. We don't want
to be restricted as such. Using the microdynamics of a quantum system as
starting point we view thermal fluctuations as statistical variations of the
coarse-grained environmental variables with which the quantum system interacts,
the exact microdynamics of the system and the environment obeying the laws
of quantum mechanics. The problem under study can thus be stated equivalently
as finding the uncertainty relation in an open quantum system.

On the third issue, in loose terms, one often identifies the high temperature
regime of a system as the classical domain. On the one hand, one often regards
the regime when thermal fluctuations begin to surpass quantum fluctuations
as the transition point from quantum to classical. On the other hand, from the
wave picture of quantum mechanics we know that a necessary condition for a
system to behave classically is that the interference terms in its wave
function have to vanish,
%
%
so that probability can be assigned to classical events [9]
%
%
or that classical decoherent histories can be well-defined [10]. This is
known as
the decoherence process. Is there any relation between these two criteria of
classicality? We show in this problem under the conditions studied that they
are indeed equivalent: The time the quantum system decoheres is also the time
when thermal fluctuation overtakes quantum fluctuations [see Eq. (27)].

However we issue a warning here that this regime should not be called
classical,
as is customary in many quantum to classical transition studies. In fact,
after the decoherence time only the first postulate of quantum statistical
mechanics (QSM) is satisfied, the system can be described by {\it
non-equilibriu
   m}
QSM. Only after the relaxation time, when the second postulate is satisfied,
can one use {\it equilibrium} QSM. Classical has still a long way to go.
It is well-known that quantum statistical effects can be important at very high
temperatures (e.g., Fermi temperature for metals). This is due to exchange
interactions of identical particles, a distinctly quantum effect. Only when
the statistical properties of fermions and bosons can be approximated by
distinguishable particles, usually at high temperatures when the Fermi-Dirac
or Bose-Einstein statistics approaches the Maxwell-Boltzmann statistics, can
the system be rightfully called {\it classical}. In this regard {\it quantum}
carries two meanings, one refers to the interference effect and the other
refers

to spin-statistics effect.


In this letter we use a simple model of a quantum open system to examine these
issues. We study the uncertainty principle at finite temperature because it is
a simple problem where these issues manifest clearly. The model we use is that
of a collection of coupled harmonic oscillators where one is distinguished as
the system of interest and the rest as bath. We use the influence functional
method [11] to incorporate the statistical effect of the bath on the system.
As the microdynamics is explicit in this approach, one can study how the result
depends on the properties of the bath and the system-bath interaction. This
mode
   l
has been studied extensively [5-6][11-13], so we won't belabor too much the
deta
   ils.


Our system is a Brownian particle with mass $M$ and natural frequency $\Omega$.
The environment is modeled by a set of $n$ harmonic oscillators with mass $m_n$
and natural frequency $\omega_n$. The particle is coupled linearly to the $n$th
oscillator with strength $C_n$. The action of the combined system and
environmen
   t
is
$$
{\eqalign{
S[x,q]
& = S[x]+S_b[q]+S_{int}[x,q] \cr
& = \int\limits_0^tds\Biggl[
    \Bigl\{{1\over 2}M\dot x^2-{1\over 2}M\Omega_0^2 x^2\Bigr\}
  + \sum_n\Bigl\{{1\over 2}m_n\dot q_n^2
  - {1\over 2}m_n\omega^2_nq_n^2 \Bigr\}
  + \sum_n\Bigl\{-C_nxq_n\Bigr\}\Biggr]\cr}}    \eqno(4)
$$
\noindent where $x$ and $q_n$ are the coordinates of the particle and the
oscillators respectively, and $\Omega_0$ is the (bare) frequency of the
particle. We are interested in how the environment affects the system in
some averaged way.  The quantity containing this information is the reduced
density matrix of the system $\rho_r(x,x')$ obtained from the full density
operator of the system and environment $\rho(x,q;x'q')$ by tracing out the
environmental degrees of freedom $ (q,q') $
$$
\rho_r(x,x',t)
=\int\limits_{-\infty}^{+\infty}dq
 \int\limits_{-\infty}^{+\infty}dq'
 \rho(x,q;x',q',t)\delta(q-q')                    \eqno(5)
$$
\noindent The reduced density matrix evolves under the action of the
propagator $J_r(x,x',t~|~x_i,x'_i,0) $ in the following way:
$$
\rho_r(x,x',t)
=\int\limits_{-\infty}^{+\infty}dx_i
 \int\limits_{-\infty}^{+\infty}dx'_i~
 J_r(x,x',t~|~x_i,x'_i,0)~\rho_r(x_i,x'_i,0~)   \eqno(6)
$$
In general, this is a very complicated expression since the evolution
operator $J_r$ depends on the initial state. If we assume that at a given
time $t=0$ the system and the environment are uncorrelated, i.e. that
$$
\hat\rho(t=0)=\hat\rho_s\times\hat\rho_e,       \eqno(7)
$$
\noindent then the evolution operator for the reduced density matrix can
be written as
$$
J(x_f,x'_f,t~|~x_i,x'_i,0)
=\int\limits_{x_i}^{x_f}Dx
 \int\limits_{x'_i}^{x'_f}Dx'~
 \exp{i\over\hbar}\Bigl\{S[x]-S[x']\Bigr\}~F[x,x']
                                                 \eqno(8)
$$
\noindent where $F[x,x']$ is the Feynman-Vernon influence functional [11].
If the environment is initially in thermal equilibrium at a temperature
$T=\beta^{-1}$ for the problem described by (4), the influence functional
can be computed exactly. The result is well known [11,12]:
$$
\eqalign{
F[x,x']=\exp\biggl\{
& -{i\over\hbar}\int\limits_0^tds_1\int\limits_0^{s_1}ds_2
   \Bigl[x(s_1)-x'(s_1)\Bigr]\eta(s_1-s_2)
   \Bigl[x(s_2)+x'(s_2)\Bigr] \cr
& -{1\over\hbar}\int\limits_0^tds_1\int\limits_0^{s_1}ds_2
   \Bigl[x(s_1)-x'(s_1)\Bigr]\nu(s_1-s_2)
   \Bigl[x(s_2)-x'(s_2)\Bigr]\biggl\} \cr}
                                                   \eqno(9)
$$
\noindent The non-local kernels $ \eta $ and $ \nu $ are defined as
$$
\eqalignno{
\nu(s)&
 =\int\limits_0^{+\infty}d\omega~I(\omega)
  \coth({{\hbar \omega} \over {2 k T}})  ~
  \cos\omega s                                      &(10a)\cr
\noalign{\hbox{and}}
\eta(s)&={d\over ds}~\gamma(s)                      &(10b)\cr
\noalign{\hbox{where}}
\gamma(s)&
=\int\limits_0^{+\infty}
d\omega~{I(\omega)\over\omega}~\cos\omega s.         &(10c)\cr}
$$
\noindent Here $I(\omega)$ is the spectral density function of the environment,
$$
I(\omega)= \sum\limits_n{\delta(\omega -\omega_n)}
           {{C^2_n}\over{2 m_n \omega_n}}.           \eqno(11)
$$
An environment is classified as ohmic $I(\omega)\sim\omega$, supra-ohmic
$I(\omega)\sim\omega^n , n>1$ or sub-ohmic $n<1$. The most studied ohmic
case corresponds to an environment which induces a dissipative force
linear in the velocity of the system. An example which we have studied
has spectral density given by [12]
$$
I(\omega)
=M\gamma_0 \omega ({\omega\over \tilde\omega})^s
e^{-{\omega^2\over\Lambda^2}}                      \eqno(12)
$$
\noindent where $\tilde\omega$ is a frequency scale usually taken to be
the cut-off frequency $\Lambda$.


The real and imaginary exponents of $F[x,x']$ are usually regarded as
responsible for dissipation
and noise respectively, thus the names dissipation and noise kernels are given
to $\eta$ and $\nu$. The most general environment would then engender nonlocal
dissipation and colored noises. We refer the reader to Ref [5] for a discussion
of the fluctuation-dissipation relation and the time scales for the relevant
processes.

The propagator has been calculated before [5,11-13]. (We use the notation of
Ref. 5)
$$
\eqalign{
J(x_f,x'_f,t~|~x_i,x'_i,0)
& = Z_0(t)  \exp{i\over\hbar} \Bigl\{
   \Bigl[\dot u_1(0)X_i+\dot u_2(0)X_f\Bigr]Y_i
  -\Bigl[\dot u_1(t)X_i+\dot u_2(t)X_f\Bigr]Y_f\Bigr\} \cr
& \times\exp{-1\over\hbar}\Bigl\{a_{11}(t)Y_i^2
  +[a_{12}(t)+a_{21}(t)]Y_iY_f
  +a_{22}(t)Y_f^2\Bigr\} \cr }                        \eqno(13)
$$
\noindent where $ X=(x+x')/2 $ and $Y=x'-x$. The elementary functions
$ u_a(s) $ are obtained as solutions of
following differential equations
$$
{d^2u_i(s)\over ds^2}
+2\int\limits_0^sds'\eta(s-s')u_i(s')
+\Omega^2u_i(s)=0                                  \eqno(14)
$$
\noindent with boundary conditions
$$
\Biggl\{
\eqalign{
& u_1(0)=1, ~~~  u_1(t)= 0 \cr
& u_2(0)=0, ~~~  u_2(t)= 1 \cr }
                                                           \eqno(15)
$$
\noindent and the $ a_{ij}(t) $ are obtained from the integral
$$
a_{ij}(t)
=\int\limits_0^tds_1\int\limits_0^{s_1}ds_2
 v_i(s_1)\nu(s_1-s_2)v_j(s_2)                       \eqno(16)
$$
\noindent where $ v_1(s)=u_2(t-s) $ and $ v_2(s)=u_1(t-s)$

We now consider a Brownian oscillator with an initial wave function
$$
\psi(x_i,0) = N_0~e^{-\frac{x_i^2}{4\sigma^2}}      \eqno(17)
$$
\noindent where $\sigma$ is the initial spread of the Gaussian packet. One
can calculate $\rho_r(x_f, x'_f, t)$ by performing the Gaussian integrals
over $x_i$ and $x'_i$ and get
$$
\rho(x_f, x'_f, t)
=\Bigl[Z_0(t) N_0^2 {\pi\over\sqrt{det {\bf H}}}\Bigr]
\exp\Bigl\{-{1\over 2} {\bf X^T}{\bf Q}^{-1} {\bf X}\Bigr\}
                                                   \eqno(18)
$$
The prefactor (terms within the square bracket, call it $\tilde N_0 (t)$)
depends only on time. Here ${\bf X}=(X,Y)^{\bf T}$ and  $Q_{ij}(t)$ is a
$2\times 2$ matrix whose elements will be given below.

To calculate the averages of observables, it is convenient to use the
Wigner function defined as
$$
W(X,p,t) = \int dY e^{{i\over \hbar} p Y} \rho (X- {Y \over 2},
             X+ {Y \over 2}, t),                         \eqno(19)
$$
%
%
%
\noindent The quantum average of an observable, e.g., $x^n$, with
respect to a pure state is given by
$$
< x^n>_0 = \int dx x^n \rho(x,x,t)
         = \int dx \int {{dp} \over {2 \pi \hbar}} x^n W (X, p, t)
                                                        \eqno(20a)
$$
\noindent and
$$
<p^n>_0 =  \int dx \int {{dp} \over {2 \pi \hbar}} p^n W (X, p, t)
                                                        \eqno(20b)
$$
\noindent Similar relations exist between $\rho_r$ and $W_r$. The averages
with respect to a mixed state now weighted by  $\rho_r$ or $W_r$ have both
quantum and thermal contributions. We get
$$
<x^2>_T = {1\over Q_{11}(t)}                           \eqno(21a)
$$
\noindent and
$$
<p^2>_T = \hbar^2 {\det Q~(t)\over Q_{11}(t)}
                                                        \eqno(21b)
$$
\noindent From them, with $ (\Delta x)_T^2=<x^2>_T-<x>_T^2 $ and
$ (\Delta p)_T^2=<p^2>_T-<p>_T^2 $,
$$
U_T (t) = (\Delta x)_T^2 (\Delta p)_T^2
        = \hbar^2 {\det Q(t) \over [Q_{11}(t)]^2 }.
        = \hbar^2 \Bigl\{ {Q_{22}(t) \over Q_{11}(t)}
        - {Q_{12}(t) \over Q_{11}(t)}
          {Q_{21}(t) \over Q_{11}(t)} \Bigr\}.
                                                        \eqno(22)
$$
\noindent Now we write out the matrix elements in (22)
$$
\eqalign{
{Q_{22}(t) \over Q_{11}(t)}
& = {1\over 4} { [\dot u_1(t)]^2 \over [\dot u_2(0)]^2 }
+ {2\sigma^2\over\hbar [\dot u_2(0)]^2} \Bigl\{
   [\dot u_1(t)]^2 a_{11}(t)
- \dot u_1(0) \dot u_1(t) [a_{12}(t)+a_{21}(t)] \cr
& + \Bigl[ {\hbar^2\over 4\sigma^4}
+ [\dot u_1(0)]^2\Bigr] a_{22}(t) \Bigr\}
+ {1\over [\dot u_2(0)]^2} \Bigl\{
  4a_{11}(t)a_{22}(t)-[a_{12}(t)+a_{21}(t)]^2 \Bigr\} \cr }
                                                        \eqno(23a)
$$
\noindent and
$$
\eqalign{
{Q_{12}(t) \over Q_{11}(t)}
= {Q_{21}(t) \over Q_{11}(t)}
& = {i\over\hbar} \sigma^2 {\dot u_2(t)\over [\dot u_2(0)]^2}
\Bigl\{ {\hbar^2\over 4\sigma^4} + [\dot u_1(0)]^2
-{\dot u_2(0)\over \dot u_2(t)} \dot u_1(0) \dot u_1(t)\Bigl\} \cr
& +{i\over \dot u_2(0)} \Bigl\{ 2 {\dot u_2(t) \over \dot u_2(0)}
a_{11}(t) + a_{12}(t) + a_{21}(t) \Bigr\} \cr }
                                                        \eqno(23b)
$$
The exact result of this expression can be obtained from solving the
equations for the $u_i(s)$ functions and the $a_{ij}(t)$
coefficients.
The results from numerical integrations are presented in Fig. 1 for
selected parameters. It is also instructional to look at some special
cases where one can find analytical results in the asymptotic regimes.

Assume an ohmic environment ($n=1$ in (12)) then
$ \gamma (t) = 2 \gamma_0 \delta (t) $ and $a_{ij}(t)$, $u_i(t)$ are
simple harmonic and exponential
functions. For weak coupling (small $\gamma_0$) and assuming a quantum
minimum-uncertainty initial state where ${\hbar\over 2\sigma^2\Omega}=1$
(also is ground state of a harmonic oscillator), we get for all temperatures,
$$
U_T(t)={\hbar^2\over 4}\Bigl[
   e^{-\gamma_0 t} + \coth({\hbar\Omega\over 2kT})
   (1- e^{-\gamma_0 t})\Bigr]^2.
                                                   \eqno(24)
$$
\noindent This is a simple, clean and intuitively clear result. We see that
there are two factors at play here: time and temperature. Time is measured
in units of the relaxation time proportional to $t_{rel}=\gamma_0^{-1}$, and
temperature is measured with reference to the ground state energy
$\hbar \Omega /2$ of the system. At $t=0$, $ U_T(0)=\hbar^2/4$,
which is the Heisenberg relation (1),
when the initial uncorrelated conditions (7) is assumed valid. At very
long time ($t>>\gamma_0^{-1}$), $ U_T(t) $ approaches $ U^{BE}_T $ as in
(3) at finite temperature, or $ U^{BM}_T $ as in (2) at high temperature.
That means the system (the Brownian particle) approaches an equilibrium
quantum statistical system. (For supraohmic bath this may not always be true).
Call $z= {\hbar \Omega\over 2kT}$. At zero temperature, $\coth z=1$ and
$U_T(t)= U_0^{QM}$ as in (1) at all times, as expected. At high temperature
($\coth z \approx 1/z$) and at short times ($t<<\gamma_0^{-1}$) this simplifies
to
$$
U_T(t) = {\hbar^2\over 4}
\Bigl[ 1+\Bigl({2kT\over \hbar\Omega}-1\Bigr)\gamma_0t
       + O(t^2) \Bigr]                                   \eqno(25)
$$

These simple  expressions are revealing in several aspects: Note that in
the expression for short time behavior (25) the first term is the ubiquitous
quantum fluctuation, the second term is the thermal contribution, which depends
on the initial spread and increases with increasing dissipation and
temperature.

The time when thermal fluctuations overtake quantum fluctuations is when the
second term in the square bracket becomes larger than unity which occurs at
(the temperature is higher than the ground state energy by assumption)
$$
t_1 = {{ \hbar \Omega_0} \over {2 \gamma_0 kT}}     \eqno(26)
$$
\noindent This is indeed  the decoherence time scale $t_{dec}$ [5-6,9],
the time when the off-diagonal components of the reduced density matrix
go to zero, and the first postulate of quantum statistical mechanics
becomes valid. The second time scale is the relaxation time scale,
$t_{rel} = \gamma_0^{-1}$ we referred to earlier, when the particle reaches
equilibrium with the environment. It is at this time that the second postulate
of quantum statistical mechanics becomes valid and equilibrium QSM can be
applied to the combined system + environment. After this, for ohmic and
subohmic

environments the uncertainty relation takes on the Bose-Einstein form (3).
At high temperatures the system reaches the Maxwell-Boltzmann limit and the
uncertainty relation takes on the classical form (2). For weak coupling and
supraohmic environments at low temperature, the highly nonlocal frequency
respon
   se
makes it difficult for the system to settle down. The decoherence time scale is
longer, and the relaxation can even be incomplete.
%
%
This is the regime where one expects to find more intricate and interesting
behavior in the interplay of quantum and thermal effects. The nonohmic results
and details of the present study are to be presented elsewhere [14].



This work is supported in part by the National Science Foundation under
grant PHY 91-19726.


\vskip 0.2in

\noindent{\bf Figure Caption}

\noindent Fig. 1.  Plotted here is the change of the uncertainty function
$U_T(t)$ in time $t$, where $ t=0 $ is the time when the system is brought
in contact with a bath at temperature $ T $. Here the parameters are
chosen to be $ \gamma_0=0.1$, $\Lambda=200$ and $T=10000K$, for which the
decoherence time is $ 5\times 10^{-4}$ and relaxation time is 10 (in units
where
$\Omega=1$).

\vfill
\eject

\noindent {\bf References}

\noindent [1] I. Bialynicki-Birula and J. Mycielski, Comm. Math. Phys. 44,
    129 (1975);     W. Beckner, Ann. Math. 102, 159 (1975).

\noindent [2] D. Deutsch, Phys. Rev. Lett. 50, 631 (1983);
    M. H. Partovi, Phys. Rev. Lett. 50, 1883 (1983).

\noindent [3] See, e.g., S. Abe and N. Suzuki, Phys. Rev. A41, 4608 (1990)

\noindent [4] See, e.g., Y. S. Kim and W. Zachary, eds {\it Squeezed State and
    Uncertainty Relation} (NASA Publication, 1992)

\noindent [5] B. L. Hu, J. P. Paz and Y. Zhang, Phys. Rev. D45, 2843 (1992)

\noindent [6] W. G. Unruh and W. H. Zurek, Phys. Rev. D40, 1071 (1989)

\noindent [7] See, e.g., K. Huang, {\it Statistical Mechanics}, 2 ed. (Wiley,
    New York, 1987)

\noindent [8] A. Mann, M. Revzen, H.Umezawa and Y. Yamanaka, Phys. Lett.
              A140, 475 (1989)

\noindent [9] W. H. Zurek, Phys. Rev. D24, 1516 (1981); D26, 1862 (1982);
    in {\it Frontiers of Nonequilibrium Statistical Physics},
    ed. G. T. Moore and M. O. Scully (Plenum, N. Y., 1986);
    E. Joos and H. D. Zeh, Z. Phys. B59, 223 (1985)

\noindent [10]M. Gell-Mann and J. B. Hartle, in {\it Complexity, Entropy
             and the Physics of Information},
             ed. W. Zurek, Vol. IX (Addison-Wesley, Reading, 1990)

\noindent [11]R. P. Feynman and F. L. Vernon, Ann. Phys. 24, 118 (1963)

\noindent [12]A. O. Caldeira and A. J. Leggett, Physica 121A, 587 (1983);
    A. J. Leggett et al, Rev. Mod. Phys. 59, 1 (1987)

\noindent [13]H. Grabert, P. Schramm, and G. -L. Ingold,
                     Phys. Rep. 168, 115 (1988)

\noindent [14]B. L. Hu, A. Raval and Y. Zhang, in preparation
\end